\documentclass[acs,twocolumn,amsmath,amssymb,reprint]{revtex4}
\usepackage{graphicx}
\usepackage{amsmath}
\usepackage{latexsym}
\usepackage{amsmath}
\usepackage{amssymb}
\usepackage{float}
\usepackage{color}
\usepackage{epstopdf}
\usepackage{hyperref}
\hypersetup{urlcolor=red,citecolor=blue}
\usepackage{bm}
\hypersetup{urlcolor=red,citecolor=blue}

\usepackage{epsfig}
\newcommand{\spthree}{$sp^3$}
\newcommand{\gowo}{G$_0$W$_0$}

\begin{document}

\title{Self-energy corrected tight binding parameters for few $p$-block semiconductors
in the hybridized atomic orbital basis constructed from first principles}
\author{Manoar Hossain, Joydeep Bhattacharjee}
\affiliation{National Institute of Science Education and Research, HBNI, Jatni,
Khurda, Bhubaneswar, 752050, Odisha, India}
\begin{abstract}

We present self-energy corrected tight-binging(TB) parameters in the basis of the directed 
hybridised atomic orbitals constructed from first principles, 
for nano-diamonds as well as bulk diamond and zinc blende structures made of elements of group 13, 14 and  15 in the 
2$p$, 3$p$ and 4$p$ blocks. 
With increasing principal quantum number of frontier orbitals, 
the lowering of self-energy corrections(SEC) to the band-gap and consequently to the dominant inter-atomic  TB parameters, 
is much faster in bulk than in nano-diamonds and hence not transferable from bulks to nano-structures. 
However, TB parameters transfered from smaller nano-diamonds to much larger ones exclusively
through mapping of neighbourhoods of atoms not limited to nearest neighbours, 
are found to render HOMO-LUMO gaps of the larger nano-diamonds with few hundreds of atoms 
in good agreement with their explicitly computed values at the DFT as well as DFT+\gowo \ levels.
TB parameters and their SEC  are found to vary significantly from 2$p$ to 3$p$ block but negligibly from 3$p$ to 4$p$,
while varying rather slowly within each block, implying the possibility of transfer of SEC across block
with increasing principal quantum number.
The demonstrated easy transferability of self-energy corrected TB parameters in the hybrid orbital basis thus 
promises computationally inexpensive 
estimation of quasi-particle electronic structure of large finite systems with thousands of atoms.


\end{abstract}
\maketitle
\section{Introduction}
\label{intro}
Starting with the advent of the quantum theory of solids \cite{bloch}, 
tight-binding(TB) frameworks  in the basis of spatially localized Wannier functions (WF) \cite{ashcroft1976solid}, 
have been envisioned as a generic approach to compute electronic structure of materials.
Slater-Koster (SK) parametrization\cite{slater1954simplified,koster1954wave} 
of elements of a TB Hamiltonian marked the beginning of reasonable band structure calculations of solids.
With the advent of the Kohn-Sham(KS) density functional theory(DFT)
\cite{hohenberg1964inhomogeneous,kohn1965self},  
the mean field approximation of KS-DFT \cite{lda1,gga2} 
and tight-binding parameters \cite{goringe1997tight,vogl1983semi,tb_ordern1} 
determined by fitting the energy bands at the DFT level,
have been the work horse for studies of ground states properties of materials with weak to modest localization of valence electrons
for about three decades now.

With our increasing capability to fabricate devices with ever shrinking size - now well within double digits in nanometers,
interest in accurate computation of physical and electronic structure of such systems, which typically consist of
thousands of atoms, led to the evolution of tight-binding frameworks for finite systems with structural inhomogeneities.
Efforts to derive TB parameters for large finite systems, particularly with \spthree \ hybridization of orbitals,
have been reported not only in the orthogonal \spthree \ basis
\cite{tserbak1993unified,sutton1988tight,goodwin1989generating,wang1993tight,mercer1994tight,kwon1994transferable}  
but also more generalizably in the non-orthogonal basis  as well 
\cite{menon1993nonorthogonal,frauenheim1995density,bernstein1997nonorthogonal} 
and used extensively in molecular dynamics simulation
\cite{wang1993tight,sankey1989ab} 
and calculation of optical properties
\cite{delerue2000excitonic,proot1992electronic,delerue1993theoretical,niquet2000method,trani2005tight}. 
Evolution of computational techniques to construct spatially localized Wannier functions (WF) \cite{marzari1997maximally,jbwan}, 
opened up the scope for construction of realistic localized orbitals from KS single particle states,
\cite{qian2008quasiatomic,mostofi2008wannier90}, 
leading to explicit computation of realistic TB parameters from first principles.

With improved spectroscopic measurements enhanced correlation due to spatial confinement near surfaces and interfaces
of nano-structures became apparent which made it imperative to take the frameworks of computation of electronic structure beyond
the mean field regime leading to formidable increase in computational complexity and cost.
While much effort in this direction has been underway in terms of improving the exchange-correlation functionals 
\cite{heyd2003hybrid} 
to include effects of many-electron interactions within the KS energy eigen-spectrum,
the GW approximation \cite{hybertsen1986electron} 
of the many body perturbation theory (MBPT) \cite{hedin1965new,hedin1970effects}  
is so far the most general abinitio framework to unmissably account for correlations starting from the KS single particle states.
However, both the approaches are computationally  expensive and almost prohibitive for systems
with thousands of atoms. 
These limitations have led to attempts to incorporate effects of correlations within the tight-binding(TB) frameworks.
Efforts reported in recent years mostly improve TB parameters \cite{qptb1,qptb7,qptb6,qptb5,qptb4} by matching QP structure
in specific systems and orbital subspaces.

In this work we calculate self-energy correction(SEC) to TB parameters from the SEC of the KS energy eigenstates at the DFT+\gowo \ level. 
With  tetrahedrally coordinated atoms exclusively considered in this work except the passivating H atoms at the surface of the nano-diamonds,
TB parameters are calculated in the orthonormal basis of Wannier functions(WF) 
constructed following a realistic template of \spthree \ hybridized atomic orbitals(HAO) obtained 
from KS single particle states of isolated atoms. 
The resultant WFs resemble hybrid atomic orbitals and orient towards directions of coordination by construction,
and thereby locked to the immediate neighbourhoods of atoms.
The aim of this work is to present a comparative analysis of TB parameters and their SEC
in such directed hybrid orbital basis beyond the nearest neighbourhood for bulk and finite
structures made of a representative variety of  2$p$, 3$p$ and $4p$ block elements which exists in diamond or zinc-blend
structures in three dimensions.
We also  demonstrate that such TB parameters can be transferred from smaller nano-diamonds to their much larger
iso-structural counterparts exclusively through mapping of neighbourhoods typically up to 2nd or third nearest neighbours,
to render HOMO-LUMO gaps of the larger systems in good agreement with their explicitly calculated values
at the DFT and DFT+\gowo \ level.

\begin{figure}[b]
\includegraphics[scale=0.27]{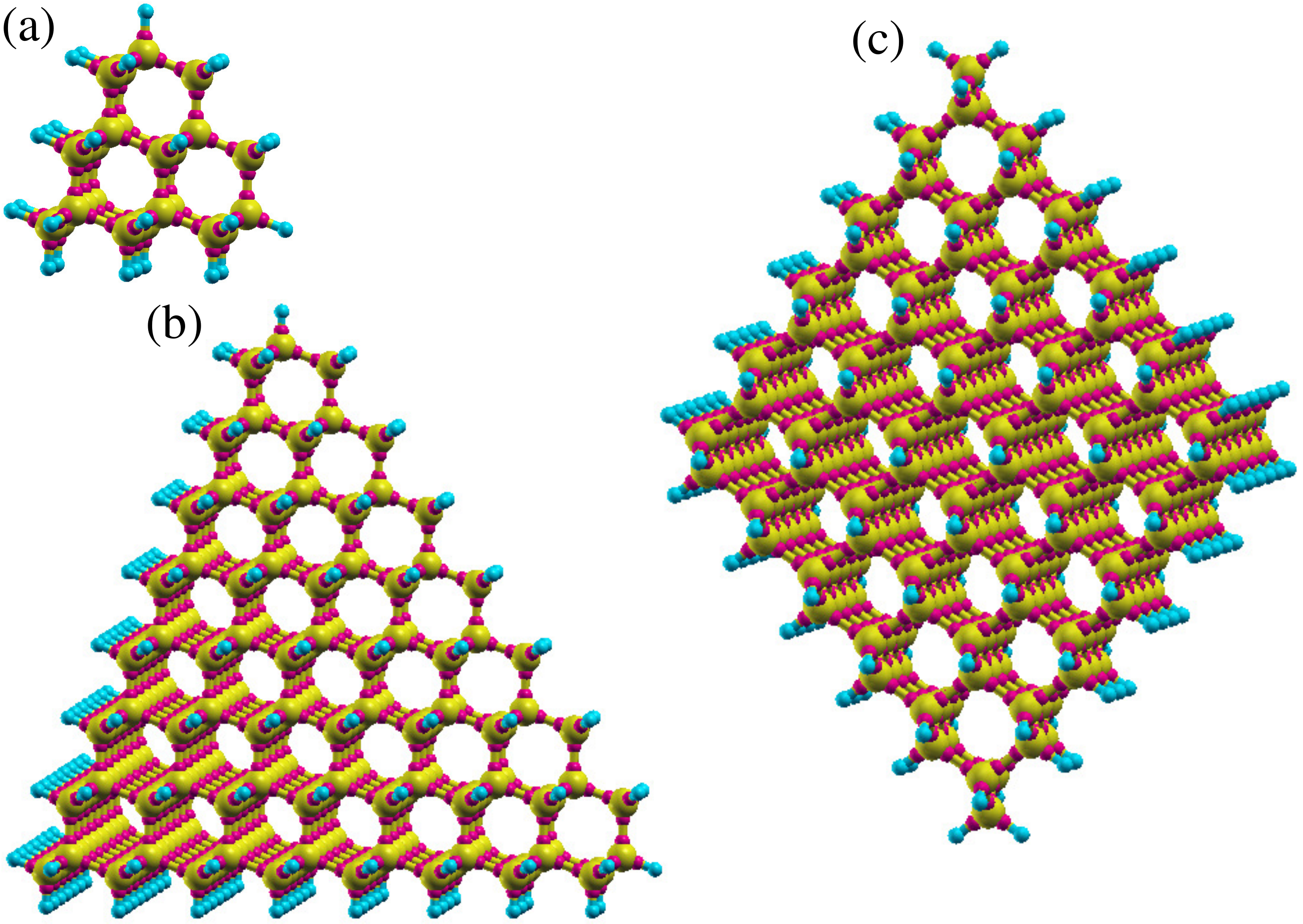}
\caption{Nano-diamonds with projected charge centres of HAOs (pink spheres):
(a) $C_{26}H_{32}$, (b) $C_{281}H_{172}$ and (c) $C_{322}H_{156}$. 
Charge centres of H coincides with H. }
\label{ndwc}
\end{figure}

\section{Methodological details}
\subsection{Construction of hybrid atomic orbitals(HAO) and their transfer to system of atoms}
Hybrid orbital basis (HAO) are constructed for isolated atoms as the approximate eigenstates
of the non-commuting set of three first moment matrices (FMM) $X,Y$ and $Z$ projected in a finite subspace of orthonormal basis states.
FMMs calculated as:
\begin{equation}
(X,Y,Z)_{ij}=\langle \phi_i \mid  (x,y,z)  \mid \phi_j \rangle. 
\label{firstmom}
\end{equation}
are maximally joint diagonalized using an iterative scheme based on the Jacobi method of matrix diagonalization
where off-diagonal elements are minimized through rotation of coordinates by a specific analytic choice of angles
as detailed in ref.\cite{hossain2021hybrid}. 
HAOs are naturally oriented towards direction of coordinations around 
atom  as per the set of basis states considered for construction of the HAOs. 
The charge centres of HAOs thus render coordination polyhedra around an atom for a given sub-shell.
For construction of the \spthree \ HAOs used in this work, 
the lowest four  KS states of a single atom are used as basis.
Although in principle, any localized atomic orbitals like the pseudo-atomic orbitals or the Slater or Gaussian 
type orbitals parametrized for the given atoms can also be used for the purpose,
it is important to have appropriate length-scale of localization of the HAOs for their correct representation 
within the KS states of the given system of atoms where the HAOs are to be used as template to construct WFs.  
Same pseudo-potential  thus preferably be used for generation of HAOs and also for calculation of electronic structure
of the given system.


Separate sets of HAOs are constructed for each types of atoms, and transfered to the corresponding atoms 
in the given system through mapping of directions
of coordination from the atoms to their nearest neighbours on to the directions of charge centres of HAOs from the 
isolated atom for which they are explicitly constructed.
Through such mapping a distribution  of projected charge centres of HAOs are generated for the entire system prior to the 
transfer of HAOs, as seen in Fig.\ref{ndwc}. 
For perfect tetrahedral coordination, as seen in the bulk diamond and zinc blende structures, the transfered HAOs
would thus retain intra-atomic orthonormality, whereas for the nano-diamonds such intra-atomic orthogonality will be
marginally compromised owing to deviations of perfect tetrahedral coordination more towards the surface. 
However this is not a  problem as such not only because the deviations are minimal but because all transfered HAOs are 
symmetrically orthonormalized during the construction of the WFs.     

\begin{figure}[t]
\includegraphics[scale=0.23]{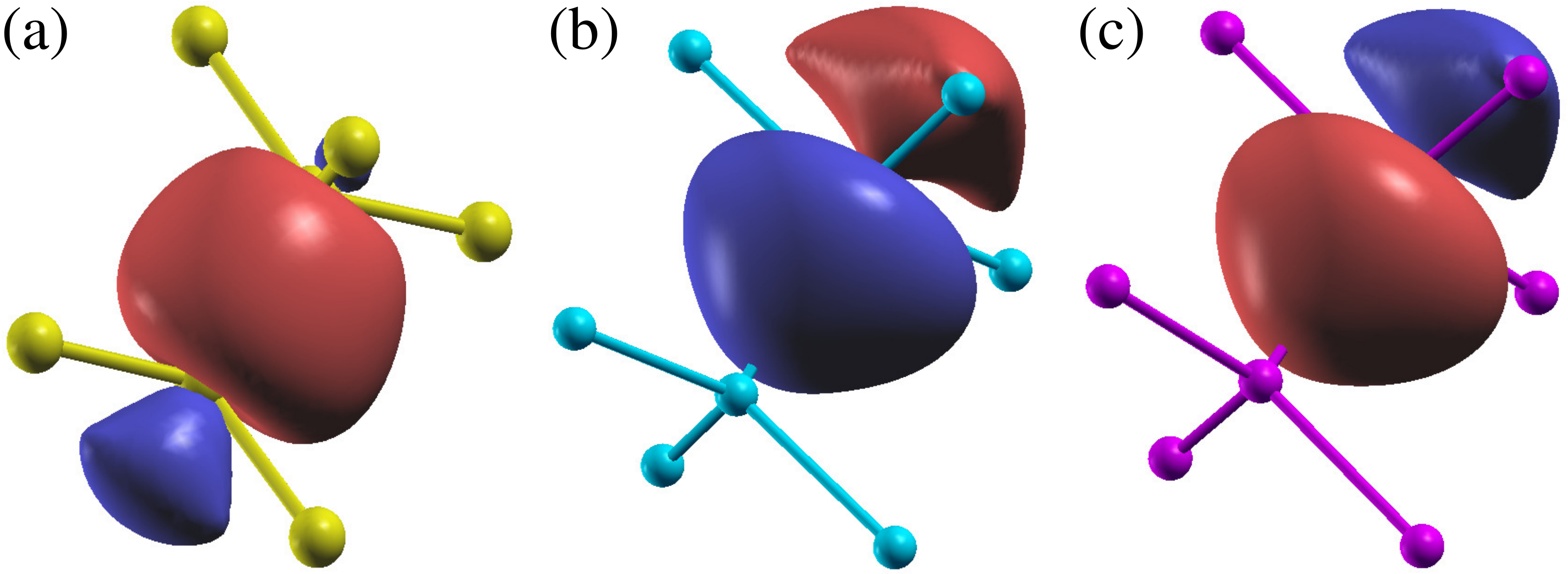}
\caption{HAWOs in bulk (a)C (b)Si and (c)Ge.}
\label{ndwf}
\end{figure}

\begin{figure*}[t]
\includegraphics[scale=0.22]{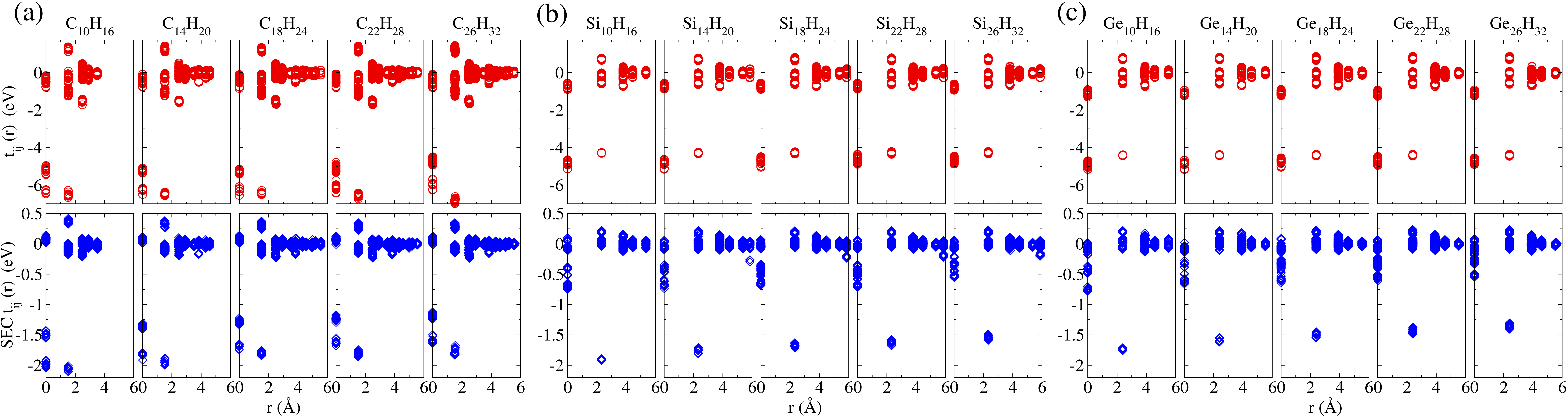}
\caption{TB and self-energy correction to TB parameters in (a) carbon, (b) silicon and (c) germanium nano-diamonds from adamantane to pentamantane.}
\label{ndtij}
\end{figure*}
\begin{figure}[b]
\includegraphics[scale=0.31]{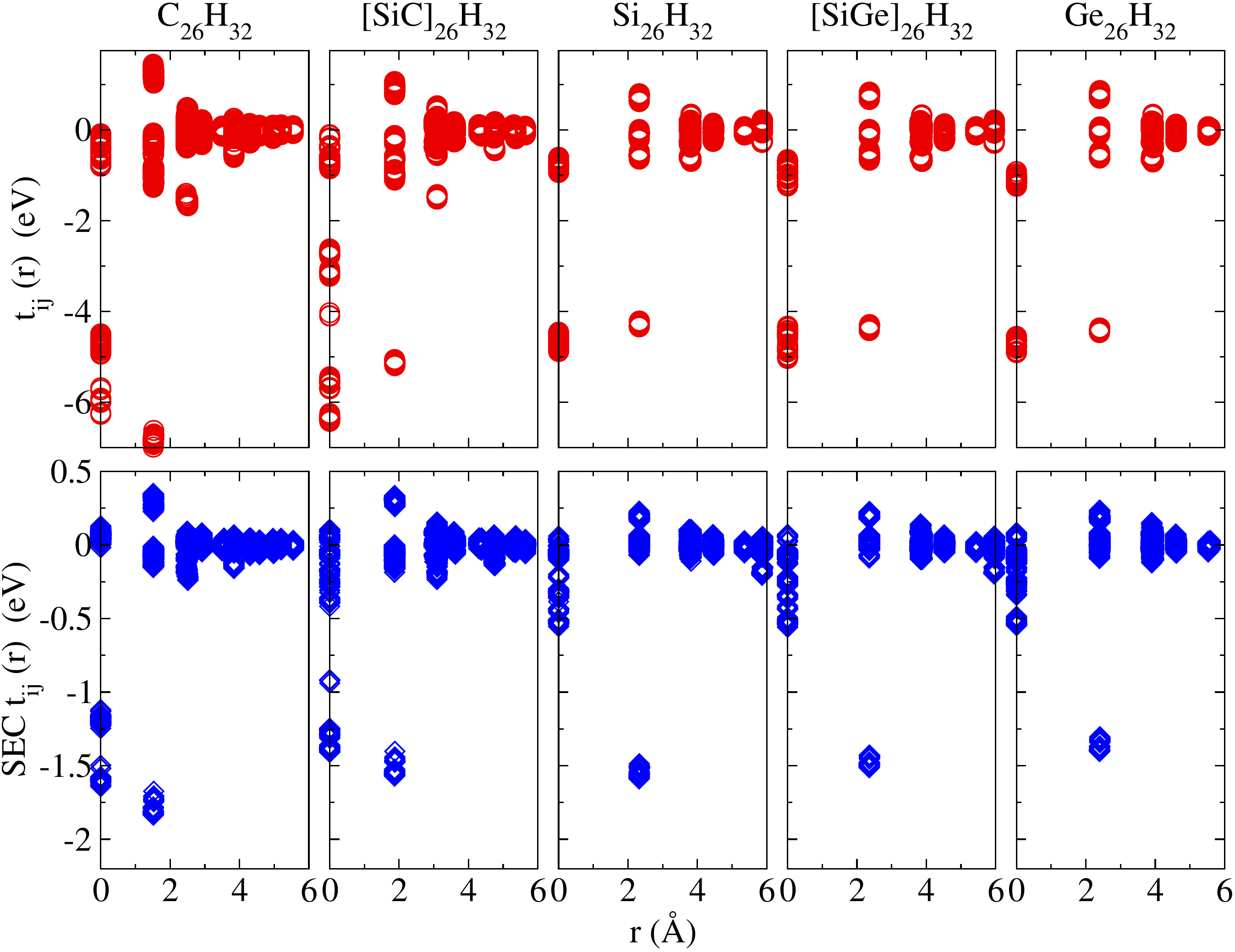}
\caption{TB and self-energy correction to TB parameters of hybrid SiC and SiGe nano-diamonds(pentamantane) along with those of C, Si and Ge nano-diamonds(pentamantane).}
\label{ndhyb}
\end{figure}
\subsection{Wannier functions based on HAOs}
\label{mhawo}
HAOs transferred from isolated atoms to the system of atoms constitute a non-orthogonal  basis primarily due to their
inter-atomic non-orthogonality.
A set of quasi-Bloch states are subsequently constructed as:  
\begin{equation}
\tilde{\psi}_{\vec{k},j}(\vec{r}) = \frac{1}{\sqrt{N}} \sum_{\vec{R}} e^{i \vec{k} \cdot \vec{R}} \phi_{\vec{R},j} (\vec{r}),
\end{equation}
where $ \phi_{\vec{R},j} (\vec{r})$ is the $j$-th HAO localized
in the unit-cell denoted by the lattice vector $\vec{R}$ which spans over $N$ unit-cells as per the the 
Born-von Karman periodicity.
The quasi-Bloch states are projected on the orthonormal Bloch states constructed from the 
KS single-particle states at all allowed $\vec{k}$:
\begin{equation}
 O_{\vec{k},m,j}= \langle \psi^{KS}_{\vec{k},m}  \mid \tilde{\psi}_{\vec{k},j} \rangle.
\label{proj}
\end{equation}   
The subspace of KS states is of size same as the number of total number of HAOs transfered to the system,
which is twice the number of occupied states in our case.  
Overlaps between the non-orthogonal quasi-Bloch states are calculated within the manifold of the considered KS states as:
\begin{equation}
S_{\vec{k},m,n}= \sum_l O^*_{\vec{k},l,m} O_{\vec{k},l,n}. 
\label{overlap}
\end{equation}   
%
\begin{figure*}[t]
\includegraphics[scale=0.41]{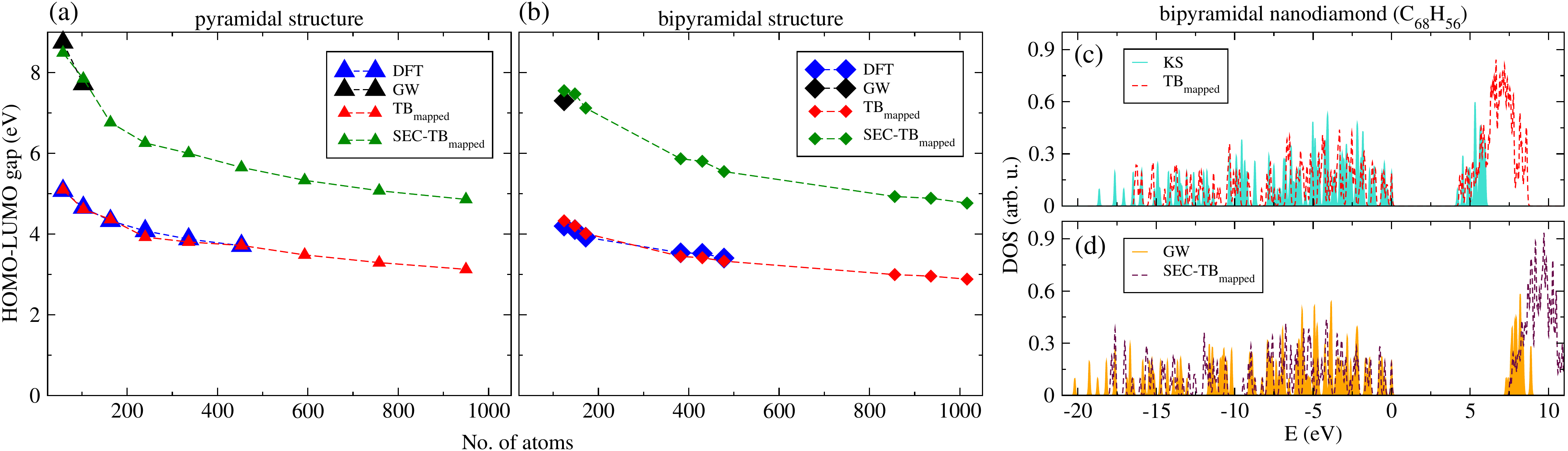}
\caption{Variation of HOMO-LUMO gap computed with TB and SEC-TB parameters mapped to un-relaxed structures from relaxed pentamantane structure, 
and the same computed from first principles at the DFT and DFT+\gowo \ level with relaxed structure, 
as function of size of carbon nano-diamonds increasing in (a) pyramidal and (b) bi-pyramidal structures. 
Comparison of DOS of C$_{68}$H$_{56}$ calculated from (c) TB mapped to  un-relaxed structures and DFT of relaxed structure,  and 
(d) SEC-TB mapped to un-relaxed structure and DFT+\gowo \ of relaxed structure .}
\label{ndmapped}
\end{figure*}
\begin{figure}[b]
\includegraphics[scale=0.335]{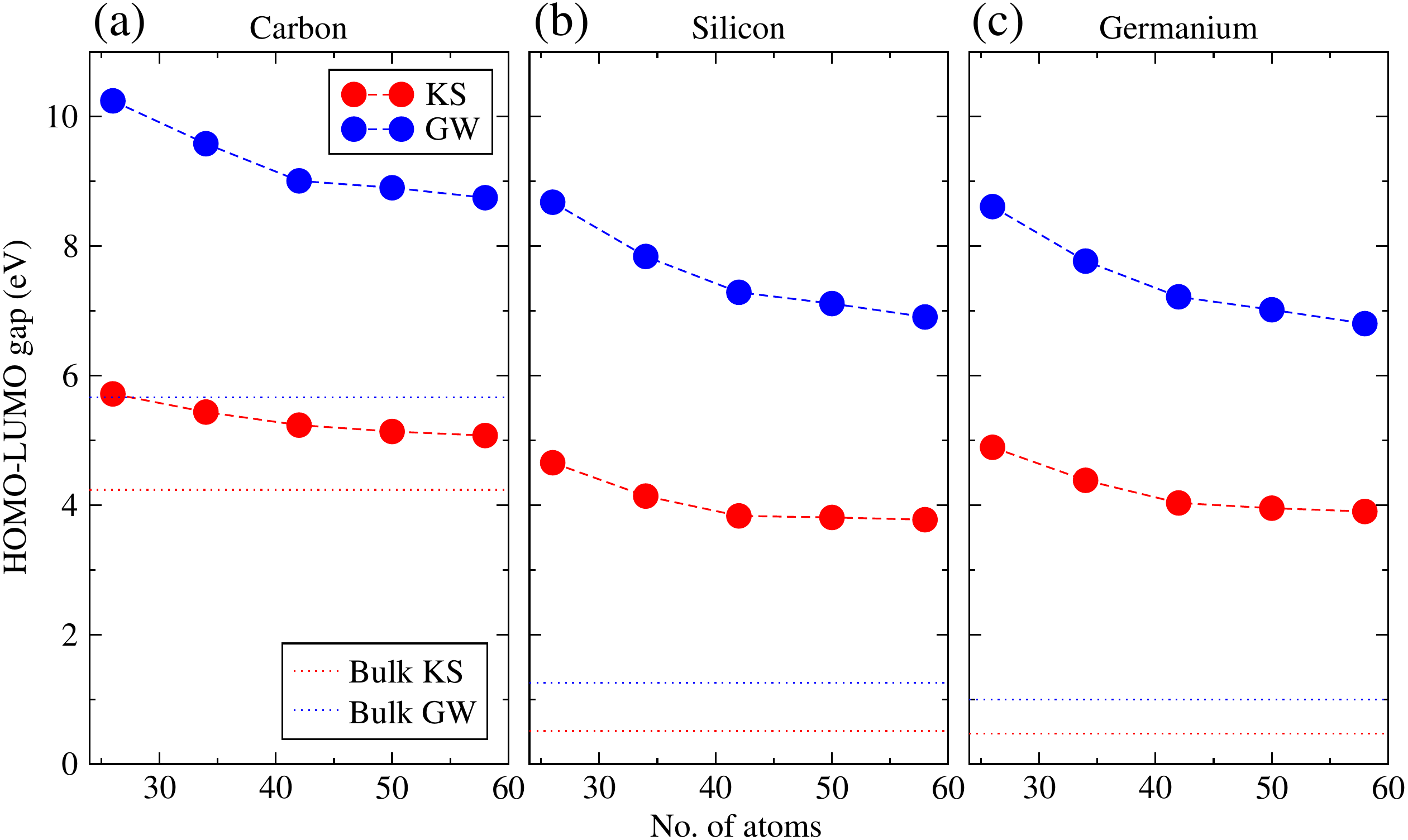}
\caption{Variation of HOMO-LUMO gap of nanodiamonds from adamantane to pentamantane made
of (a) carbon, (b) silicon and (c) germanium.  Bulk band gaps are shown in dotted lines.}
\label{gap}
\end{figure}
Representability of an HAO within the set of KS states considered, 
is ensured by setting a lower cutoff on individual $S_{\vec{k},n,n}$ values for all HAO index $n$, which we typically set at 0.85.
In all the systems considered in this work, such a cutoff is found to be well satisfied for all HAOs 
within the set of KS states whose count is set to the total number of HAOs of all the atoms in the given system. 
Using the L\"{o}wdin symmetric orthogonalization \cite{lowdin1950non} scheme a new set of orthonormal Bloch states 
are constructed from the KS single particle states as:
\begin{equation}
\Psi_{\vec{k},n}(\vec{r}) = \sum_m S^{-\frac{1}{2}}_{\vec{k},m,n} \sum_l O_{\vec{k},l,m} \psi^{KS}_{\vec{k},l}(\vec{r}),
\label{awobf}
\end{equation}
where $l$ spans over the KS states and the sum over $m$ ensures orthonormalization.
Finally a orthonormal set of localized Wannier functions referred here onwards in this paper 
as the  hybrid atomic Wannier orbitals (HAWO), are constructed as:
\begin{equation}
\Phi_{\vec{R'},j}(\vec{r}) = \frac{1}{\sqrt{N}} \sum_{\vec{k}}  e^{-i \vec{k}\cdot \vec{R'}} \Psi_{\vec{k},j}(\vec{r}).
\label{awo}
\end{equation}   
%
The resultant HAWO resemble the corresponding HAOs $[\left\{ \phi_{\vec{R},j} (\vec{r})  \right\}]$ as seen in Fig.\ref{ndwf}.
Notably, the number of KS states which can be considered for construction of HAWOs are not limited by the total number of
HAOs transfered to the system. In fact the localization of the HAWOs increase with increasing size of the KS subspace used.
However in such a case the matrix $(OS^{-\frac{1}{2}})$ would become semi-unitary.
%
TB parameters are computed in the HAWO basis from energetics of KS single particle states as:
\begin{eqnarray}
& &t_{\vec{R'},\vec{R},i,j} = \langle \Phi_{\vec{R'},i} \mid H^{KS} \mid \Phi_{\vec{R},j} \rangle \nonumber \\
&=&\sum_{\vec{k}}e^{i\vec{k}.(\vec{R'}-\vec{R})}\sum_l (OS^{-\frac{1}{2}})^*_{li} (OS^{-\frac{1}{2}})_{lj} E^{KS}_{\vec{k},l}.
\label{hop}
\end{eqnarray}
Notably, although TB parameters have been derived from first principles based on  the
maximally localized Wannier function \cite{marzari2012maximally,lee2005band,atomicorb2,calzolari2004ab,franchini2012maximally,jung2013tight} 
or atomic orbitals \cite{atomicorb1,qian2010calculating,d2016accurate,agapito2016accurate} constructed from KS states over the years,
attempts to calculate TB parameters in the basis of directed hybrid orbital  has been primarily  limited so far 
to analytical models\cite{yue2017thermal,popov2019deductive}. 
Self-energy corrected TB parameters (SEC-TB) $\left\{t^{QP}_{\vec{R'},\vec{R},i,j} \right\}$ in the HAWO basis are calculated 
by substituting $E^{KS}_{\vec{k},n}$ in Eqn.(\ref{hop}) by quasi-particle energies $E^{QP}_{\vec{k},n}$ obtained at
the $G_0W_0$ level of the GW approximation of MBPT\cite{hedin1965new,hybertsen1986electron}.
Efforts have been reported in recent years on incorporating SEC in TB parameters 
computed in terms of the MLWFs\cite{qptb3,qptb5,qptb6}.
%

\begin{figure*}[t]
\includegraphics[scale=0.34]{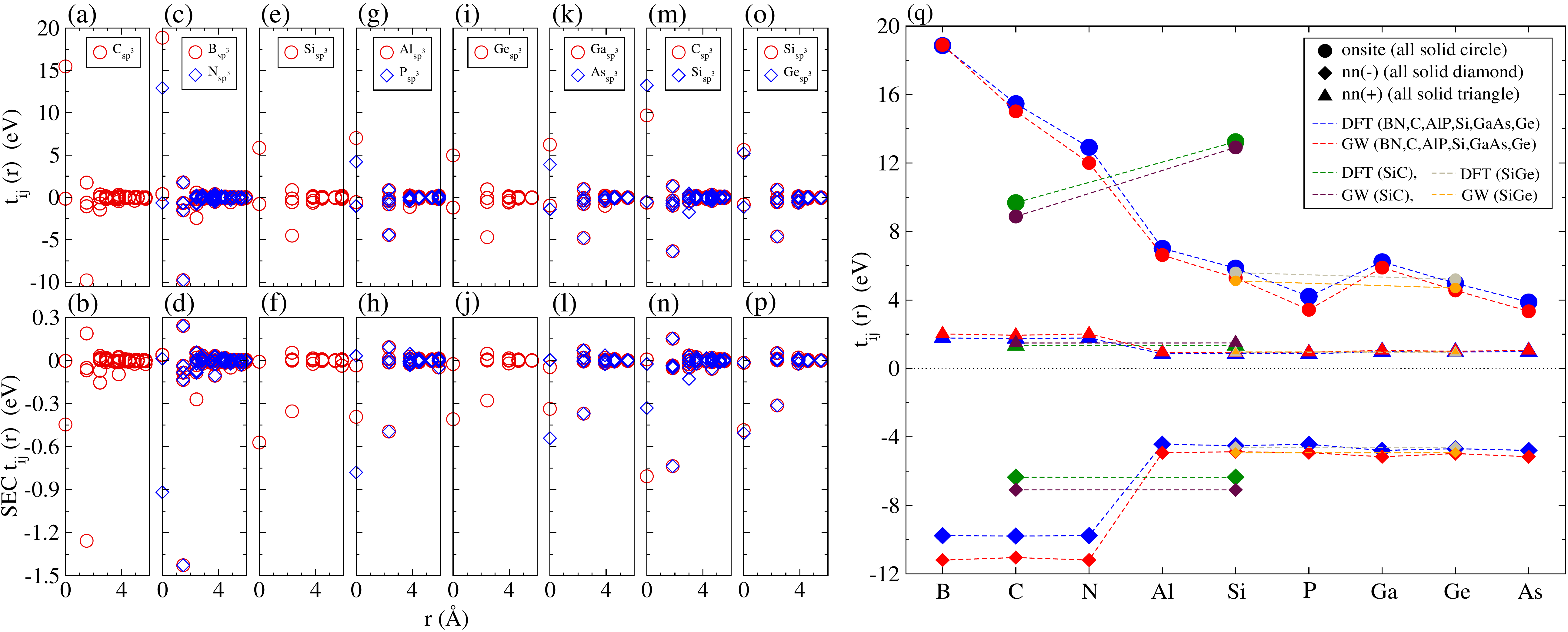}
\caption{(a-p) TB parameters and their self-energy corrections, and (q) summary of onsite and dominant inter-atomic terms,
 for  2$p$, 3$p$ and 4$p$ block elements in diamond and zinc blende structures.}
\label{bulktij}
\end{figure*}
\subsection{Bottom-up mapping of TB parameters}
\label{mapr2t}
The orientation of HAWOs being locked to the local atomic neighbourhood of each atom,
the multi-orbital TB parameters derive in HAWO basis are easily mappable from one system
to another with matching atomic neighbourhoods.
Transferability of the TB parameters in terms of reproducibility of band-gaps and band-width within acceptable ranges
of deviation from their direct estimates from first principles, is demonstrate by transferring TB parameters 
from smaller reference systems to structurally similar larger target systems.
%
%
%
Transfer of TB parameters is done in two steps.\\ 
First all individual pairs of atoms of the target system beyond the  nearest neighbourhood,
are mapped on to matching pairs of atoms in the reference system through comparison of a collection of structural parameters like spatial 
separation, angles and dihedral angles subtended by the nearest neighbours and a measure ($\zeta$)  of proximity of atoms with 
respect to surfaces, interfaces and inhomogeneities defined as:
\begin{equation}
\zeta_i = \sum^{N_i}_j Z_j w(r_{i,j})
\label{mapparam}
\end{equation}   
$N_i$ being the number of neighbours of the $i$-th atom within a chosen cutoff radius, 
and $w$ being a weight factor which is a function of the distance $r_{i,j}$ of the $j$-th neighbour of the $i$-th atom,
and $Z_i$ being a characteristic number to be associated with each type of atom, for example, the atomic weight.

For mapping of orbital pairs from reference to target, a projected 
map [Fig.\ref{ndwc}(b-c)] 
of locations of charge centres  
of HAOs for the target system is generated first and then 
among the mapped pair of atoms, pair of HAOs of system are mapped to pair of HAOs of the reference
through mapping of their respective charge centres much like the mapping of atom pairs. 
%
%
%
The weight factor $w$ is typically chosen to be 1.0 within half of the cutoff radius and smoothly reduced
to zero at the cutoff radius using a cosine function.  
Cutoff radius is chosen based on the size of the reference system, since it should neither be too large for variations in $Z_i$ 
to average out, nor should it be too small such that $\zeta$ varies abruptly only at the structural inhomogeneities.
Cutoff radius and $Z_i$ values should thus be so chosen that $\zeta$ values can effectively differentiate between atoms 
at the interior of a finite system from those at the surfaces in both reference and target systems such that relaxed structures of 
target systems are not required to adequately transfer effects of variation in bond lengths near the surfaces from reference to target systems.

\section{Computational details}
All the ground state geometries as well as ground state electronic structures are calculated 
using the QUANTUM ESPRESSO (QE) code \cite{giannozzi2009quantum} 
which is a plane
wave based implementation of density functional theory (DFT). 
BFGS scheme has been used to obtain the relaxed structures.
For bulk systems variable cell relaxation has been performed to optimize lattice parameters and ionic positions.
A separation of at least 10 \AA\ between periodic images has been ensured for all the finite systems.
The KS ground state properties are calculated within the Perdew-Zunger approximation of the  
exchange-correlation functional implemented in a norm-conserving pseudo-potential.
Plane wave basis with kinetic energy cutoff in excess of 60 Rydberg 
along with a 15x15x15 Monkhorst-Pack grid of k-points have been used for all the bulk systems. 
The self-energy correction to KS energy eigenvalues are calculated at the \gowo \ level of 
GW approximation of the many-body perturbation theory (MBPT) 
implemented in the BerkeleyGW (BGW) code\cite{deslippe2012berkeleygw}. 
To calculate the static dielectric matrix and extend to the finite
frequencies the generalized plasmon-pole model \cite{hybertsen1986electron} 
is used.
For convergence of dielectric function as well as the self-energy term
we used in excess of 250(5000) bands 
for bulks(nano-diamonds). To get a well converged band-gap for bulk
structures we used a finner k-mesh of 30x30x30 grid. 
Finally, in-house implementation interfaced with the QE code is used for generation of HAOs, HAWOs 
from KS states, calculation of TB parameters in the HAWO basis, 
and mapping of TB parameters from smaller reference systems to larger target systems.
%

\section{Results and discussion}
Nano-diamonds starting with adamantane followed by 
diamantane, triamantane, tetramantane 
with increasing number of atoms up to pentamantane have 
been considered for explicit calculation of TB parameters at the DFT level and their further refinement due to SEC 
of KS energy eigenvalues at the \gowo \ level.
As evident in Fig.\ref{ndtij} the dominant nearest neighbour(nn) TB parameters as well as those beyond the nearest neighbourhood,
remains effectively invariant with respect to system size, implying transferability of TB parameters 
across length-scales of nano-diamonds at the DFT level as we demonstrate in this paper. 
Magnitudes of TB parameters in Si and Ge are similar but vary significantly from that of C, which is consistent 
with variation in their inter-atomic separations. 
Notably however, the SEC to the dominant TB parameters reduces slowly with increasing system size in a similar rate
for C, Si and Ge based nano-diamonds, although the magnitude of correction itself
reduces significantly with increasing principal quantum number owing to increased de-localization.
The slow variation of SEC across system size also suggests possibility of SEC corrected TB parameters to be 
transferable as well to a workable degree.

Fig.\ref{ndhyb} clearly suggests that TB parameters of either Si or Ge can be used for SiGe nano-diamond as well, 
whereas for the SiC nano-diamonds the dominant interatomic TB parameters are effectively the
average of that of the C and Si based nano-diamonds.
These observations hint at increased transferability of SEC across elements in a block with increasing quantum number of 
frontier orbitals. 
The reorganization of the on-site terms of SiC nano-diamond  compared to those made of  Si or C can be attributed to the
hetero-polarity of the Si-C nn bond due to difference of electro-negativities of Si and C,
whereas electro-negativities of Si and Ge are similar and less than that of C.

To study transferability of TB parameters from smaller nano-diamonds to their larger counterparts we have mapped TB
parameters from pentamentane to larger pyramidal and bi-pyramidal 
carbon based nano-diamonds with close to 1000 atoms.
We have considered un-relaxed coordinates of the larger nano-diamonds and mapped the TB parameters calculated for relaxed structure 
of pentamentane, using $\zeta$ values calculated with a cutoff radius considered up to the next nearest neighbour.
As evident in Fig.\ref{ndmapped}(a,b), the match of the DFT HOMO-LUMO gap with that calculated from the transferred TB parameters from pentamentane
is found to be quite exact for both pyramidal as well as bi-pyramidal nano-diamonds up to about 500 atoms,
which is almost about ten times escalation of size of the target system compared to that of the reference system.  
Notably, for all the larger nano-diamonds the DFT HOMO-LUMO gaps were calculated explicitly with well relaxed structures, 
whereas the structures used to transfer TB parameters from pentamentane were completely un-relaxed. 
Therefore the match of the exact DFT HOMO-LUMO gaps 
with that evaluated with TB parameters transferred to an un-relaxed structure point to the efficacy of $\zeta$ in differentiating atoms 
as per their proximity to the surface in both reference and target systems so that the transferred TB parameters
have the correct variation from surface to interior.
The match of the quasi-particle (QP) HOMO-LUMO gap is also found quite satisfactory up to more than two time escalation of system size
beyond pentamentane, till which G$_0$W$_0$ calculations could be reliably performed with computational facilities at our end. 
QP HOMO-LUMO gap data available in the literature \cite{raty2005optical,raty2003quantum,drummond2005electron,sasagawa2008route} 
allows us to compare for a bit further. 
These results thus confirms transferability of TB parameters with and without SEC over substantial escalation of system size.   



\begin{figure}[t]
\includegraphics[scale=0.31]{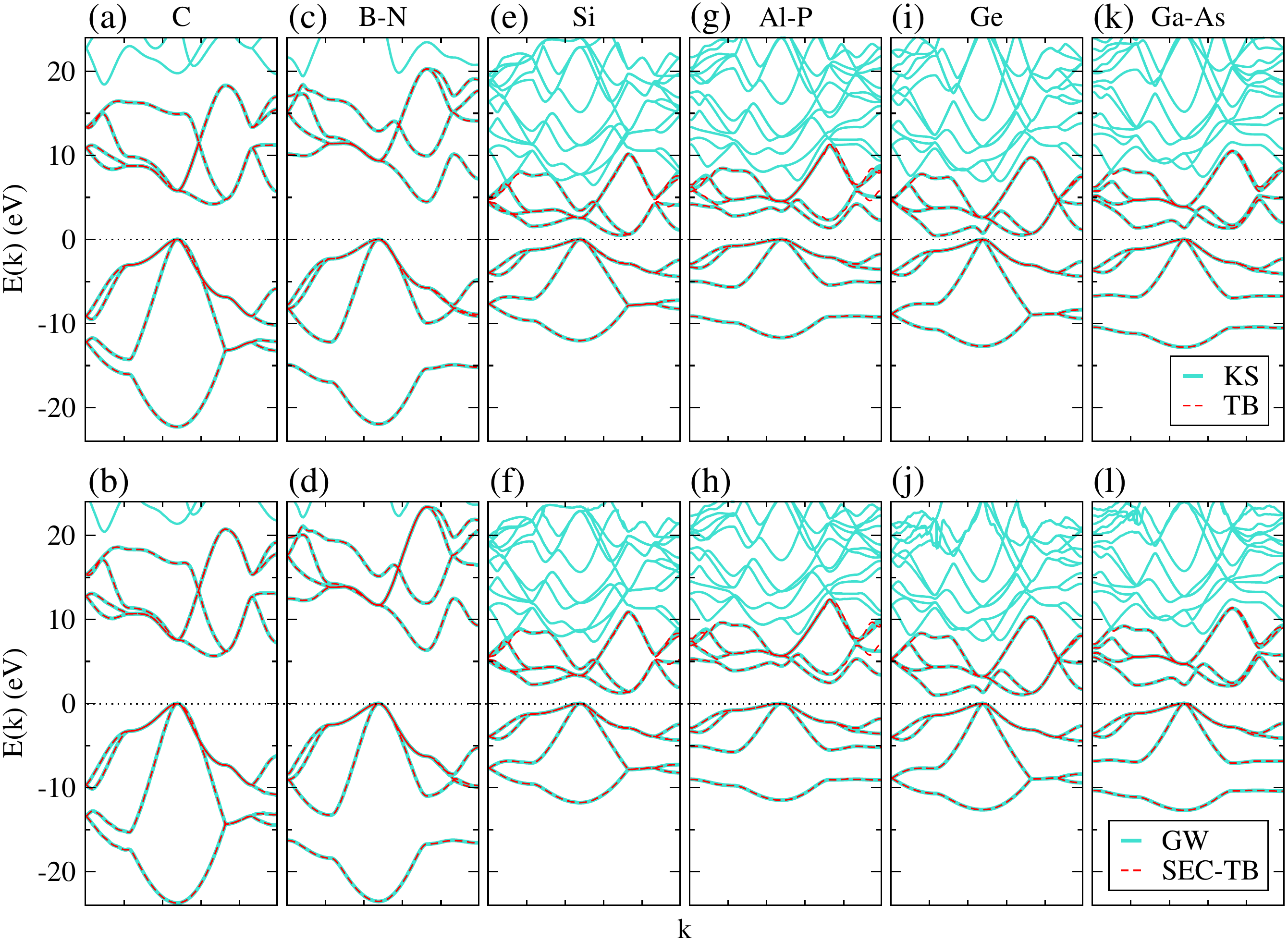}
\caption{Comparison of band-structures computed from first principles with that obtained with TB parameters derived from first principles at the 
DFT (upper panel)  and DFT+\gowo (lower panel)  levels for 2$p$, 3$p$ and 4$p$ block elements.}
\label{bulkbs}
\end{figure}

Interatomic TB parameters for bulk Si and Ge in diamond structures are consistent with those in nano-diamonds 
notwithstanding the substantial drop of band-gap in bulk to about 0.5 eV(1 eV) from HOMO-LUMO gap around 4 eV(7 eV) in nano-diamonds 
at the DFT(DFT+\gowo\ ) level as shown in Fig.\ref{gap}(b,c). 
However for bulk C in diamond structure the dominant nn hopping term strengthens substantially to -10 eV (Fig.\ref{bulktij}(a))
from -6.5 eV (Fig.\ref{ndtij}(a) red plots) in nano-diamonds, although the band-gap in bulk is comparable to that of HOMO-LUMO gap of nano-diamonds,
which are about 4.2 eV and 5 eV respectively  at the DFT level.
The reason of this contrasting trend for C compared to those for Si and Ge can be traced to
one obvious fact that \spthree \ HAWOs in Si and Ge are more de-localized than that of C owing to increase in principal quantum number,
but in addition to that, the C-C nn separation clearly decreases in bulk from nano-diamonds by about 2\% to 3 \%, 
as has also been reported \cite{saani2007comparison} 
earlier, whereas the  Si-Si and Ge-Ge nn separations remain largely same in bulk and nano-diamonds.
Furthermore, as evident in Fig.\ref{ndwf}, the major lobe (positive one in this case) of the \spthree \ HAWOs of C is stretched towards
the nearest C compared to the HAWOs  of Si and Ge. 
Overlap of such \spthree \ HAWOs of C would thus increase appreciably with reduction in C-C nn separation from nano-diamonds 
to bulk, leading to the increase in magnitude of nn hopping term. Thus the TB parameters are not transferable from bulk to
nano-diamond for C even at the DFT level. However it may still be feasible for Si and Ge. 

Consistent with the localized nature of HAWOs, SEC is higher for C based bulk and nano-diamonds compared to their 
Si and Ge based counterparts. However the relative variation of localization of HAWOs of C, Si and Ge are much dominated over by 
the localization imposed by the finiteness of nano-diamonds, 
leading to comparable values of SEC of band-gap of C, Si and Ge based nano-diamonds ranging from about 4 eV to 3 eV leading
to comparable SEC correction to nn hopping term ranging from -1.75 eV to -1.3 eV.
From nano-diamonds to bulk the SEC to band gap and thereby to the nn hopping term reduces consistently on account
of de-localization. Thus the self-energy corrected TB parameters are not transferable from bulk to nano-diamonds even for Si and Ge.

As evident in Fig.\ref{bulktij}(a,c,e,g,i,k), the nn hopping parameter is almost exactly same for diamond structures of group 14 (C,Si,Ge) 
and zinc blende structures made of groups 13 and 15 (BN,AlP,GaAs) within each block, whereas the SEC to the nn
(Fig.\ref{bulktij}(b,d,f,h,j,l)) hopping are consistently more for zinc blend structures 
than the diamond structures in each block owing likely to the enhanced localization of the bonding orbitals in the 
heteropolar B-N, Al-P and Ga-As nn bonds than in the C-C, Si-Si, and Ge-Ge bonds.   
Fig.\ref{bulktij}(q) summarizes the variation of the onsite terms and the major inter-atomic hopping parameters
at the DFT and DFT+\gowo \ level.
As evident in Fig.\ref{bulktij}(m,o,n,p,q), for zinc blend structure of SiC(SiGe) the nn hopping and it's SEC 
both are average of their counterparts in diamond structures of Si and C(Ge).
Fig.\ref{bulktij}(q) clearly suggests the stepped evolution of TB parameters and SEC from 2$p$ to 3$p$ and 4$p$ blocks and the similarity in
the later two.  
Band structures obtained with TB parameters shown in Fig.\ref{bulkbs} are plotted  along with DFT band structure. 
Notably, for the exact reproduction of band structure the TB parameters have to be obtained not only in a dense enough grid
of $\vec{k}$ but also using a KS band subspace of size exactly 
same as the number of HAOs transfered for the entire unit-cell in order to maintain the unitary nature  of the matrix $(OS^{-\frac{1}{2}})$. 
If we use a larger KS subspace then the reproduction energy bands will become increasingly inexact with increasing energy.


\section{Conclusions}
For a representative variety of $p$ block elements, we have presented TB parameters calculated at the level of DFT and DFT+\gowo \
in the orthonormal hybridized atomic Wannier orbitals(HAWO) basis constituted of the KS single particles and are
directed towards coordination by construction.
We present TB parameters for nano-diamonds made of C, Si, Ge, SiC and SiGe and their bulks in diamond or zinc blende structures,
and also of bulk BN, AlP and GaAs in the three consecutive $p$ blocks.
Transferability of inter-atomic TB parameters between bulk structures to nano-diamonds is generally poor and worsens with 
self-energy correction(SEC).
However among nano-diamonds, the inter-atomic TB parameters at the DFT level are found to remains effectively unaltered with increasing size, 
implying robust transferability of the TB parameters from smaller to larger nano-diamonds as demonstrated. 
Slow reduction of SEC to TB parameters with increasing system size  
also implies good transferability of self-energy corrected TB parameters across nano-diamonds. 
Transfer of TB parameters are performed exclusively by mapping up to second or third nearest neighbourhoods between the reference and the target systems, 
and thus does not necessitates relaxed geometries of the target systems as long as all neighbourhoods could be mapped. 
Similarity of TB parameters and their self-energy corrections across 3$p$ and 4$p$ blocks compared to that in the 2$p$ block hints 
at the possible transferability of SEC across blocks with increasing principal quantum number.

\section{Acknowledgments}
Computations have been performed in computing clusters supported by the Nanomission  of the Dept. of Sci.\& Tech. and Dept. of Atomic Energy of the Govt. of India.

\bibliographystyle{unsrt}
\bibliography{references}

\end{document}